# Calcul à la rupture en présence d'un écoulement : formulation cinématique avec un champ de pression approché

Alain Corfdir

*CERMES, ENPC-LCPC, Institut Navier, 6 et 8 avenue Blaise Pascal, Cité Descartes, Champs sur Marne, 77455 Marne-la-Vallée cedex 2, France*

Résumé

Nous cherchons ici à utiliser la méthode cinématique du calcul à la rupture dans le cas d'un milieu poreux soumis à un écoulement avec ou sans surface libre sans connaître la solution exacte du champ de pression. La méthode proposée ici repose sur l'utilisation de champs de pression approchés par défaut. Nous montrerons comment sous certaines conditions portant sur le critère de résistance et sur le champ de vitesse utilisé, l'utilisation de tels champs de pression approchés permet d'obtenir une condition nécessaire de stabilité sans avoir à déterminer exactement l'écoulement.

Abstract

**Yield design formulation for porous media subjected to flow, using approximate pressure field**. We attempt here to use the kinematic method of yield design in the case of a porous medium subjected to flow (with or without free surface), without looking for the exact solution of the pressure field. The method proposed here is based on the use of approximate pressure fields. In this paper, we show how, under different conditions concerning the yield criterion and the velocity field, the use of such approximate fields allows to obtain a necessary condition for stability without having to find the real pressure field.

*Abridged English version*

We attempt here to use the kinematic method of yield design [1, 2] in the case of a porous medium subjected to flow (with or without free surface), without looking for the exact solution of the pressure field. This problem widens a study dedicated to yield design in saturated porous medium without fluid flow calculation [3]. We first recall the formulation of yield design for a porous medium. Assuming some hypotheses on the yield criterion, we obtain a simple equation for the maximal resistant power. Finally, the use of specific approximate pressure fields allows to obtain a necessary condition for stability without having to determine the real pressure field.

We consider a porous medium $\Omega$ (Figure 1). The balance equation for porous media using the total stress tensor $\underline{\underline{\sigma}}$ is first recalled (1). Using any kinematically admissible virtual velocity field, we write (2).
Tensile stresses are counted positively. We assume that the domain of resistance G' is defined in terms of effective stresses (3). Furthermore, we assume that the criterion satisfies the condition (4). This condition is satisfied by usual criteria such as Tresca's, Mises', Coulomb's, Drucker-Prager's (Figure 2), but not by a bounded criterion.
Using (5) and bounding the power of internal stresses from above, we get (6), which is a necessary condition for stability according to yield design theory. This condition can be written using the maximal resistant power (7) and the density of maximal resistant power for G' (8). All relevant velocity fields verify the condition (9).

The kinematic method is based on the comparison of the maximal resistant power and the power of external forces. We therefore need to bound the maximal resistant power from above and the power of external forces from below. We assume that condition (4) is satisfied. We then show how it is possible to obtain a rigorous necessary condition of stability, knowing only an approximate pressure field p' bounding the real pressure field from below (10).

We consider the case of a flow without free surface. We recall the following theorem [1]: if two systems 1, 2 differ only by the resistance domain G and if $G_1(x) \subset G_2(x)$ everywhere, the potential stability domain $K_1$ is included in $K_2$. Assuming (3) and (4), this theorem can be applied to two systems differing only by internal pressure p such that $p_1 \geq p_2$ everywhere. So, one can deduce a necessary condition of stability by applying the kinematic method with an approximate pressure field p'≤p. This result allows to obtain an alternative to the method proposed in [3].

We consider now a flow with a free surface S (figure 3). For this second case, we use a method quite similar to the one used for the first case. There is, however, an additional difficulty: the power of external forces depends on the approximate pressure field used. Therefore, it is necessary to make sure that the power of external forces calculated with the approximate pressure field is lower than the one calculated with the real pressure field.

One denotes by χ the characteristic function of the part of the medium where p>0. Also p' denotes an approximate pressure field and χ' is the characteristic function of the part of the medium where p'>0. Assuming that $p \geq p'$, we get inequality (11) for χ and χ'.
One denotes by $\gamma_{sat}$ the specific weight of the saturated part of the medium, the rest of the medium being dry (specific weight $\gamma_d$), neglecting so the transition between the dry zone and the saturated zone. For the sake of simplicity, only velocity fields satisfying condition (12) are considered. An upper bound of the maximal resistant power (14) is then obtained. Splitting the external forces power into a surface part and a volume part (15) (16), one may conclude using (12) that a lower bound for the power of external forces (17) has been found. We finally prove the necessary stability condition (13) using only an approximate pressure field. The method can also be used for generalised effective stresses (19).

In this paper, we show how to use kinematic method for a porous medium subjected to flow without calculating the exact pressure field. To achieve this, we suppose that the resistance criterion is given in terms of effective stresses and is such that, if an effective stress state lies inside the resistance domain and if a hydrostatic pressure is added to it, the resulting stress state also lies inside the resistance domain. Then, a necessary condition for stability is obtained, using approximate pressure fields bounding the true pressure field from below and velocity fields such that the power of the weight is positive outside the part of the medium saturated for the approximate pressure field. Finally, we hope that this work will contribute to widen the cases where yield design allows, according to Salençon, to obtain results with rough methods, which reveal to be very useful in engineering practice.

1. Introduction

Nous cherchons à mettre en œuvre la méthode cinématique du calcul à la rupture [1,2] dans le cas d'un milieu poreux sans connaître la solution exacte du champ de pression. Cette problématique étend aux écoulements à surface libre une étude consacrée au calcul à la rupture en milieux poreux saturé [3]. Nous rappellerons d'abord la formulation du problème de calcul à la rupture en milieux poreux. Sous certaines hypothèses sur le critère de résistance, nous donnerons ensuite une expression simple de la puissance résistante maximale en fonction de la pression interstitielle. Enfin, nous montrerons comment utiliser la méthode cinématique en utilisant une solution approchée du champ de pression minorant le champ de pression réel pour le cas d'un écoulement sans surface libre puis avec surface libre.

2. Formulation du problème

2.1 Obtention de la condition cinématique pour un milieu poreux.

On considère le milieu poreux $\Omega$ constitué d'un squelette solide et d'un espace poreux. Nous supposerons que tout point de cet espace poreux est ou saturé de liquide ou vide. Le système mécanique considéré comprend à la fois le squelette et l'éventuel fluide inclus dans l'espace poreux. Nous notons $\underline{f}$ la densité volumique des forces extérieures. Cette densité est la somme des forces extérieures au système s'appliquant à la phase solide et à l'éventuelle phase liquide. Le chargement mécanique est défini par cette densité volumique $\underline{f}$ et par les conditions aux limites qui sont données par la densité surfacique de forces extérieures $\underline{F}$ (somme des forces extérieures surfaciques appliquées à chacune des phases) sur une partie $\partial\Omega_{\underline{F}}$ du bord et par un champ de vitesse $\underline{U}$ sur $\partial\Omega_{\underline{U}}$, complémentaire de $\partial\Omega_{\underline{F}}$. Plus généralement, les conditions aux limites peuvent être définies par la donnée, en tout point du bord, de la composante de la vitesse ou de la force extérieure pour chaque direction d'un repère orthogonal [1].

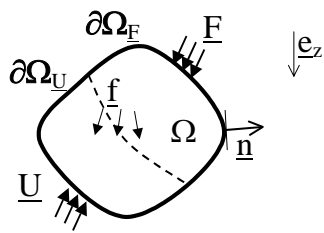

**Figure 1Définition du chargement mécanique**
Figure1 Definition of loading conditions

Si on suppose que les efforts qui s'exercent sur tout sous système peuvent être représentés par la densité volumique de forces extérieures $\underline{f}$ et par une densité surfacique de forces s'exerçant sur la frontière du sous-système, alors l'application du principe des puissances virtuelles permet d'établir l'existence d'un tenseur $\underline{\underline{\sigma}}$ décrivant les efforts intérieurs (voir par exemple [4]). Ce tenseur est le tenseur des contraintes totales et représente les efforts intérieurs sur les deux phases. Si on néglige l'accélération des particules fluides, l'équation d'équilibre s'écrit de la même manière que l'espace poreux soit vide ou saturé de liquide:

$$\text{div}(\underline{\underline{\sigma}}) + \underline{f} = 0 \qquad (1)$$

Selon l'approche statique du calcul à la rupture, un chargement est potentiellement supportable si on peut trouver un champ de contraintes $\underline{\underline{\sigma}}$ vérifiant (1) et les conditions aux limites en contraintes et satisfaisant en tout point le critère de résistance. La méthode cinématique du calcul à la rupture permet de trouver une condition nécessaire pour qu'un chargement soit potentiellement supportable. On note $\underline{\underline{\hat{d}}}$ le tenseur des vitesses de déformation virtuel, $\Sigma$ l'ensemble des surfaces de discontinuité de $\underline{\hat{U}}$, $\underline{v}$ la normale à $\Sigma$ et $[\![\underline{\hat{U}}]\!]$ le saut de $\underline{\hat{U}}$. Le principe des puissances virtuelles permet d'écrire pour tout champ statiquement admissible $\underline{\underline{\sigma}}$ et tout champ cinématiquement admissible $\underline{\hat{U}}$ :

$$\int_\Omega \underline{\underline{\sigma}} : \underline{\underline{\hat{d}}} \, dV + \int_\Sigma [\![\underline{\hat{U}}]\!].\underline{\underline{\sigma}}.\underline{v} \, dS = \int_\Omega \underline{f}.\underline{\hat{U}}. \, dV + \int_{\partial\Omega} \underline{\hat{U}}.\underline{\underline{\sigma}}.\underline{n} \, dS \qquad (2)$$

Si les conditions de chargement (forces extérieures volumiques, forces extérieures surfaciques sur $\partial\Omega_F$, vitesse sur $\partial\Omega_U$) dépendent linéairement de n paramètres de chargement, on peut écrire le membre de droite sous la forme d'un produit $\underline{Q}.\underline{\dot{q}}$, $\underline{Q}$ étant le vecteur chargement et $\underline{\dot{q}}$ le vecteur vitesse de déformation [1]. On est donc dans une situation semblable à celle d'un milieu continu classique ; toutefois, cela demande que l'éventuelle surface libre soit connue exactement puisque le poids volumique est différent entre la zone sèche et la zone saturée. Dans la suite, nous n'envisagerons que le cas où la vitesse sur le bord est nulle quand elle est imposée et où les forces extérieures volumiques se limitent au poids qu'on note $\gamma_{sat}\underline{e}_z$ quand l'espace poreux est saturé de fluide et $\gamma_d\underline{e}_z$ quand il est vide.

2.2 Introduction d'hypothèses sur le critère de résistance
On adopte la convention de la mécanique des milieux continus pour les champs de contraintes : les tractions sont comptées positivement. Un champ de pression uniforme s'écrit $-p\underline{\underline{1}}$. On se limite au cas où le critère de résistance peut être ramené à une condition portant uniquement sur les contraintes effectives. On définit alors un domaine de résistance G' en contraintes effectives.

$$\underline{\underline{\sigma}}' = \underline{\underline{\sigma}} + p\underline{\underline{1}} \in G' \qquad (3)$$

C'est par l'intermédiaire de ce critère en contraintes effectives que la pression interstitielle p intervient dans notre démarche. On fait l'hypothèse complémentaire suivante sur le domaine G' :

$$\underline{\underline{\sigma}}' \in G' \Rightarrow (\underline{\underline{\sigma}}' - \lambda \underline{\underline{1}}) \in G', \forall \lambda \geq 0 \tag{4}$$

Cette condition (4) est satisfaite pour les critères G' usuels isotropes comme les critères de Tresca, de Mises, de Coulomb, de Drucker-Prager (Figure 2). En revanche, elle n'est pas satisfaite pour des critères bornés. Elle joue un rôle central dans toute la suite.

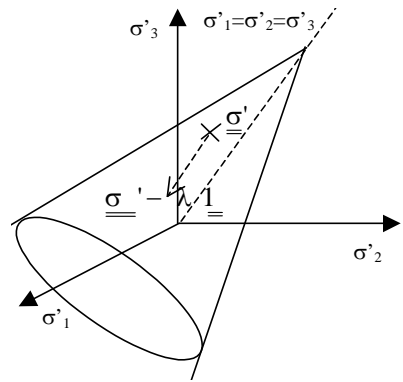

**Figure 2 Interprétation géométrique de la condition (4) dans le cas du critère de Drucker-Prager**

Figure 2 Geometrical interpretation of condition (4) for Drucker-Prager's criterion

2.3. Conséquence des hypothèses relatives au critère de résistance

On reprend la démarche de l'approche cinématique du calcul à la rupture. On note $P_e(\underline{\hat{U}})$ la puissance des efforts extérieurs. Alors grâce à (2), on a pour tout champ de vitesse virtuel $\underline{\hat{U}}$ cinématiquement admissible :

$$P_e(\underline{\hat{U}}) = \int_\Omega \underline{\underline{\sigma}} : \underline{\underline{\hat{d}}} \, dV + \int_\Sigma [[\underline{\hat{U}}]].\underline{\underline{\sigma}}.\underline{v} \, dS \tag{5}$$

En majorant le terme $\int_\Omega \underline{\underline{\sigma}} : \underline{\underline{\hat{d}}} \, dV$ par $\int_\Omega \sup\{\underline{\underline{\sigma}} : \underline{\underline{\hat{d}}} | \underline{\underline{\sigma}}' \in G'\} dV$ et $\int_\Sigma [[\underline{\hat{U}}]].\underline{\underline{\sigma}}.\underline{v} \, dS$ par $\int_\Sigma \sup\{[[\underline{\hat{U}}]].\underline{\underline{\sigma}}.\underline{v} | \underline{\underline{\sigma}}' \in G'\} dS$, on obtient une majoration de la puissance des efforts extérieurs :

$$P_e(\underline{\hat{U}}) \leq \int_\Omega \sup\{\underline{\underline{\sigma}} : \underline{\underline{\hat{d}}} | \underline{\underline{\sigma}}' \in G'\} dV + \int_\Sigma \sup\{[[\underline{\hat{U}}]].\underline{\underline{\sigma}}.\underline{v} | \underline{\underline{\sigma}}' \in G'\} dS \tag{6}$$

On appelle puissance résistante maximale et on note $P_{rm}$ le membre de droite de l'inégalité (6). En utilisant (4) on obtient pour $P_{rm}$ :

$$P_{rm}(p, \underline{\hat{U}}) = \int_\Omega \pi_{G'}(\underline{\underline{\hat{d}}}) \, dV - \int_\Omega p \, tr(\underline{\underline{\hat{d}}}) \, dV + \int_\Sigma \pi_{G'}(\underline{v}, [\![\underline{\hat{U}}]\!]) \, dS - \int_\Sigma p [\![\underline{\hat{U}}]\!].\underline{v} \, dS \qquad (7)$$

Avec $\pi_{G'}$ la densité de puissance résistante maximale volumique ou surfacique pour le critère G' qui est définie par :

$$\pi_{G'}(\underline{\underline{\hat{d}}}) = \sup\{\underline{\underline{\sigma'}} : \underline{\underline{\hat{d}}} / \underline{\underline{\sigma'}} \in G'\} \; ; \; \pi_{G'}(\underline{v}, [\![\underline{\hat{U}}]\!]) = \sup\{[\![\underline{\hat{U}}]\!].\underline{\underline{\sigma'}}.\underline{v} | \underline{\underline{\sigma'}} \in G'\} \qquad (8)$$

Les champs pertinents, c'est à dire ceux qui conduisent à une majoration effective dans (6) sont tels que $\pi_{G'}(\underline{\underline{\hat{d}}}) < \infty$ et $\pi_{G'}(\underline{v}, [\![\underline{\hat{U}}]\!]) < \infty$. Compte tenu de l'hypothèse (4), les champs pertinents vérifient (9) :

$$tr(\underline{\underline{\hat{d}}}) \geq 0 \; ; \; [\![\underline{\hat{U}}]\!].\underline{v} \geq 0 \qquad (9)$$

On retrouve là une propriété partagée par les critères de résistance usuels cités précédemment.

3. Utilisation de champs de pression approchés minorant le champ de pression réel

Nous allons voir maintenant comment obtenir une condition nécessaire de stabilité en connaissant seulement un champ de pression approché. Nous présenterons d'abord le cas d'un écoulement sans surface libre où l'espace poreux est saturé en tout point par le fluide puis le cas d'un écoulement à surface libre. Nous supposons dans toute la suite que la porosité et les poids volumiques sont constants, indépendants notamment de l'état de contrainte.

On désigne par p le champ de pression. Pour un écoulement permanent, compte tenu des hypothèses ci-dessus, le champ p est déterminé par la perméabilité du milieu, la masse volumique du fluide, les conditions en flux et en pression sur des parties complémentaires du bord du système (voir par exemple [4]). On désigne par p' un champ de pression approché par défaut. On supposera que ce champ est partout positif ou nul et est continu. On pose donc :

$$p \geq p' \geq 0 \qquad (10)$$

Si l'écoulement est permanent et sans surface libre, sous certaines conditions portant sur le champ de pression approché p', on peut vérifier que p' satisfait (10) sans connaître p en appliquant le principe du maximum (voir par exemple [5]). Le cas le plus simple est celui où les conditions aux limites hydrauliques sont données partout en pression, où la perméabilité du milieu est homogène et isotrope et où $\Omega$ est suffisamment régulier. Alors si p≥p' sur le bord et si p' satisfait la loi de Darcy et l'équation de conservation de la masse fluide, alors p et p' sont des fonctions harmoniques et on a p≥p' partout sur $\Omega$. Dans le cas d'un écoulement avec surface libre, il est plus délicat de vérifier p'<p sans connaître p [6].

3.1 Cas d'un écoulement sans surface libre

Dans le cas d'un écoulement sans surface libre, nous nous intéressons aux modes de chargement à nombre fini de paramètres. Nous pouvons alors utiliser le théorème suivant [1] : si deux systèmes mécaniques géométriquement identiques soumis au même mode de chargement sont tels qu'en tout point le domaine de résistance du matériau du premier est inclus dans le domaine du second, le domaine de stabilité potentielle du premier est inclus dans le domaine de stabilité potentielle du second. Donc, si deux systèmes 1 et 2 diffèrent seulement par la pression interstitielle et si en tout point $p_1(x)>p_2(x)$, alors, sous (3) et (4), le domaine de résistance $G_1(x)=G'(x)-p_1\underline{1}$ est inclus dans $G_2(x)=G'(x)-p_2\underline{1}$ et le domaine de stabilité potentiel $K_1$ est inclus dans $K_2$.

La méthode cinématique appliquée avec un champ de pression approché par défaut conduit donc à une condition nécessaire de stabilité pour le système réel. On peut de même utiliser la méthode statique en utilisant un champ de pression approché par excès pour obtenir un condition suffisante de stabilité potentielle [3]. L'utilisation conjointe des deux méthodes peut permettre d'évaluer l'erreur globale due, d'une part, à l'utilisation de champs de vitesse et de champs statiques qui ne correspondent pas à la solution exacte du problème de calcul à la rupture et due, d'autre part, à l'utilisation de champs de pression approchés.

3.2 Cas d'un écoulement avec surface libre

On se place maintenant dans le cas d'un écoulement avec surface libre S (Figure 3). Il y a toutefois une difficulté supplémentaire par rapport au cas de l'écoulement sans surface libre : le poids volumique en un point varie selon qu'il y a ou non présence de fluide dans l'espace poreux en ce point. La puissance des forces extérieures va donc dépendre de la position de la surface libre comme on le verra sur les équations (15), (16). Il faudra donc s'assurer que la puissance des forces extérieures calculées avec le champ de pression approché est inférieure à celle calculée avec le champ de pression réel.

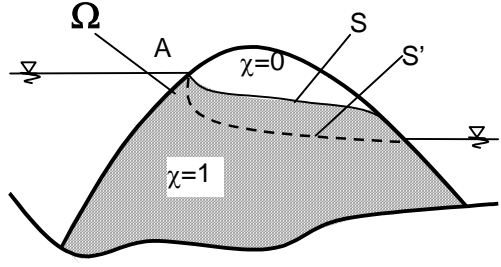

**Figure 3 Ecoulement avec surface libre**

Figure 3 Flow with free surface

On désigne par χ la fonction caractéristique de l'ensemble des points avec p>0. On considère encore un champ de pression approché par défaut p' et on désigne par χ' la fonction caractéristique de l'ensemble des points avec p'>0 et par S' la surface libre correspondante. On suppose toujours que p≥p' et on en déduit :

$$\chi \geq \chi' \tag{11}$$

On rappelle que l'on note $\gamma_{sat}\underline{e}_z$ le poids volumique du milieu dans la zone saturée (χ=1) et $\gamma_d \underline{e}_z$ dans le reste du milieu, négligeant ainsi la transition entre la zone sèche et la zone saturée. On se limitera aux champs cinématiques $\hat{\underline{U}}$ tels que la quantité $\underline{e}_z.\hat{\underline{U}}$ soit positive en tout point de la zone supposée sèche (telle que χ'=0) :

$$\underline{e}_z.\hat{\underline{U}} \geq 0 \text{ si } \chi'= 0 \tag{12}$$

Nous allons maintenant vérifier que la condition cinématique obtenue avec un champ de pression p' satisfaisant (10) et un champ de vitesse satisfaisant (12) est bien une condition nécessaire de stabilité pour le problème avec le champ de pression réel (inconnu) p. Pour cela, nous allons montrer que :

$$P_{rm}(p,\hat{\underline{U}}) \geq P_e(\chi,\hat{\underline{U}}) \Rightarrow P_{rm}(p',\hat{\underline{U}}) \geq P_e(\chi',\hat{\underline{U}}) \tag{13}$$

Par (7), (9) et (10), on obtient pour les champs $\hat{\underline{U}}$ pertinents :

$$P_{rm}(p',\hat{\underline{U}}) \geq P_{rm}(p,\hat{\underline{U}}) \tag{14}$$

Examinons la puissance des forces extérieures : elle se décompose en un terme de bord sur $\partial\Omega$ et un terme de volume sur $\Omega$. Ce terme de volume ne met en jeu que le poids total et on a :

$$P_e(\chi, \underline{\hat{U}}) = \int_{\partial\Omega} \underline{F}.\underline{\hat{U}} \, dS + \int_{\Omega} (\chi\gamma_{sat} + (1-\chi)\gamma_d)\underline{e}_z.\underline{\hat{U}} \, dV \qquad (15)$$

$$P_e(\chi', \underline{\hat{U}}) = \int_{\partial\Omega} \underline{F}.\underline{\hat{U}} \, dS + \int_{\Omega} (\chi'\gamma_{sat} + (1-\chi')\gamma_d)\underline{e}_z.\underline{\hat{U}} \, dV \qquad (16)$$

Notons que dans (16) la force surfacique extérieure $\underline{F}$ est celle du problème réel, indépendamment du choix du champ de pression approché. Si la force $\underline{F}$ n'est pas donnée sur une partie du bord de $\Omega$, alors la condition à la limite y est donnée en vitesse. Comme on se limite aux cas où cette vitesse sur le bord est nulle quand elle est imposée (2.1), la contribution correspondante de $\underline{F}.\underline{\hat{U}}$ est nulle dans (15) comme dans (16). Nous pouvons donc évaluer la différence entre $P_e(\chi)$ et $P_e(\chi')$ :

$$P_e(\chi, \underline{\hat{U}}) - P_e(\chi', \underline{\hat{U}}) = \int_{\Omega} (\chi - \chi').(\gamma_{sat} - \gamma_d)\underline{e}_z.\underline{\hat{U}} \, dV \qquad (17)$$

L'intégrande est non nulle seulement si $\chi=1$, $\chi'=0$. Et alors, compte tenu de l'hypothèse (12), et de $\gamma_{sat} \geq \gamma_d$, on vérifie que l'on a minoré la puissance des forces extérieures :

$$P_e(\chi, \underline{\hat{U}}) \geq P_e(\chi', \underline{\hat{U}}) \qquad (18)$$

En utilisant (14) et (18), l'implication (13) est démontrée. La condition cinématique obtenue avec un champ de pression approché par valeurs inférieures et un champ de vitesse virtuel vérifiant (12) est donc bien une condition nécessaire de stabilité qui conduit à une borne supérieure du chargement maximal supportable. Un exemple de mise en œuvre pratique est donné dans [6]. Notons que la démarche peut être adaptée au cas où la condition (12) n'est pas vérifiée partout. En remplaçant dans (16) $\gamma_d$ par $\gamma_{sat}$ pour les points tels que $\chi'=0$ et $\underline{e}_z.\underline{\hat{U}} < 0$ on obtient une nouvelle expression $P'_e(\chi', \underline{\hat{U}})$ qui satisfait $P_e(\chi, \underline{\hat{U}}) \geq P'_e(\chi', \underline{\hat{U}})$, inégalité analogue à (18) et on a l'analogue de (13) en y remplaçant $P_e(\chi', \underline{\hat{U}})$ par $P'_e(\chi', \underline{\hat{U}})$.

4. Extension à des critères en contraintes effectives généralisées

Nous avons supposé que le critère pouvait s'écrire en contraintes effectives. On peut envisager l'extension au cas où le critère s'exprime en contraintes effectives généralisées $\underline{\underline{\sigma}}''$ (19). Cette situation a été étudiée par divers auteurs ; on peut se référer notamment à [7] pour une démarche basée sur l'homogénéisation.

$$\underline{\underline{\sigma}}'' = \underline{\underline{\sigma}} + bp\underline{\underline{1}} \text{ avec } 1 > b > 0 \qquad (19)$$

On suppose que le critère de résistance en contraintes effectives généralisées G'' vérifie (4). On a alors l'analogue de (7) :

$$P_{rm}(p,\underline{\hat{U}}) = \int_\Omega \pi_{G''}(\underline{\hat{d}}) \, dV - b\int_\Omega p tr(\underline{\hat{d}}) \, dV + \int_\Sigma \pi_{G''}(\underline{v},[[\underline{\hat{U}}]]) \, dS - b\int_\Sigma p[[\underline{\hat{U}}]].\underline{v} \, dS \qquad (20)$$

On peut conclure comme précédemment que la puissance résistante maximale calculée avec un champ de pression 0≤p'≤p est un minorant de la pression résistante maximale. Le reste du raisonnement est identique.

5. Conclusion

Nous avons montré comment utiliser la méthode cinématique pour un milieu poreux soumis à un écoulement avec ou sans surface libre et obtenir ainsi une borne cinématique rigoureuse sans connaître le champ de pression du fluide. Pour cela, nous avons supposé que le critère de résistance était défini en termes de contraintes effectives et qu'il vérifiait de plus la propriété suivante : si un état de contraintes effectives appartient au domaine de résistance, un autre état obtenu en lui superposant un état de pression hydrostatique appartient encore au domine de résistance. Pour simplifier, nous avons utilisé des champs de vitesse tels que la densité de puissance du poids dans la zone supposée sèche (dans le champ de pression approché utilisé) était positive ou nulle, la méthode pouvant toutefois être adaptée si cette condition n'était pas satisfaite. Sous ces hypothèses, l'utilisation d'un champ de pression approché minorant le champ de pression réel permet d'obtenir une condition nécessaire de stabilité. Pour conclure, nous espérons que cette méthode peut contribuer à étendre les cas où le calcul à la rupture a déjà permis, suivant l'expression de Salençon, l'obtention « à coup de serpe » de résultats qui se sont révélés très utiles dans la pratique pour l'ingénieur.

Remerciements

Références